\documentclass{ws-procs9x6}
\newcommand{\calr}{\mathcal{R}}
\newcommand{\calp}{\mathcal{P}}
\newcommand{\ch}{\mathcal{H}}

\begin{document}

\title{Trans-Planckian Physics and Inflationary Cosmology}

\author{Robert H. Brandenberger}

\address{Physics Dept., Brown University, \\
Providence, R.I. 02912, USA\\ 
E-mail: rhb@het.brown.edu}


\maketitle

\abstracts{
Due to the quasi-exponential redshifting which occurs during
an inflationary period in the very early Universe, wavelengths
which at the present time correspond to cosmological lengths
are in general sub-Planckian during the early stages of inflation.
This talk discusses two approaches to addressing this issue
which both indicate that the standard predictions of inflationary
cosmology - made using classical general relativity and
weakly coupled scalar matter field theory - are not robust
against changes in the physics on trans-Planckian scales. One
approach makes use of modified dispersion relations for a usual
free field scalar matter theory, the other uses some properties
of space-time noncommutativity - a feature expected in string
theory. Thus, it is possible that cosmological observations may be
used as a window to explore trans-Planckian physics. We also
speculate on a novel way of obtaining inflation based on
modified dispersion relations for ordinary radiation.}

\section{Introduction}

Inflationary cosmology \cite{Guth} is an elegant paradigm of
the very early Universe which solves several conceptual problems
of standard cosmology and leads to a predictive theory of the
origin of cosmological fluctuations. The basic idea of inflation
is to replace the time line of Standard Big Bang (SBB) cosmology
(a late time phase of matter domination preceded by a period
in which radiation is dominant and which begins with a cosmological
singularity) by a modified time line for which during some time
interval $I = [t_i, t_R]$ - long before the time of
nucleosynthesis - the Universe is accelerating,
often involving quasi-exponential expansion of the scale factor $a(t)$
(to set our notation, we will be writing the space-time metric for
our homogeneous and isotropic background cosmology in the form
\begin{equation}
ds^2 \, = \, dt^2 - a(t)^2[dx^2 + dy^2 + dz^2] \, ,
\end{equation}
where $t$ denotes physical time and $x, y, z$ are the three Euclidean
comoving spatial coordinates - we are neglecting the spatial curvature).
The time $t_i$ stands for the beginning of the inflationary phase,
the time $t_R$ stands for the end, the time of ``reheating''.

Inflationary cosmology solves several of the conceptual problems of
SBB cosmology. In particular, it resolves the homogeneity problem,
the inability of SBB cosmology to address the reason for the near
isotropy of the cosmic microwave background (CMB), it explains the
spatial flatness of the Universe, and it provides the first ever
mechanism of cosmological structure formation based on causal physics.

\begin{figure}[th]
%\epsfxsize=10cm   %width of figure - will enlarge/reduce the figures
%\epsfbox{fig3.eps}
%\figurebox{2cm}{3cm}{} %to have a box alone 
\centerline{\epsfxsize=3.5in\epsfbox{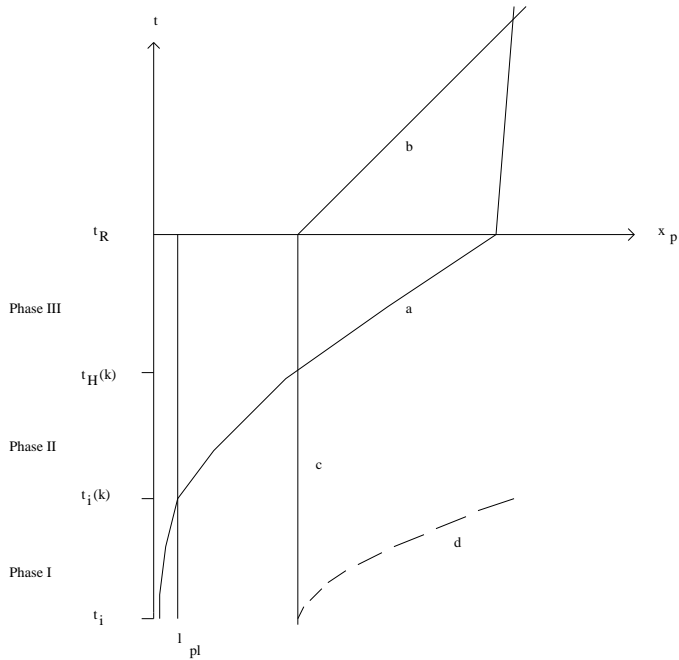}}   
\caption{Space-time diagram (physical distance vs. time)
showing how inflationary cosmology
provides a causal mechanism for producing cosmological fluctuations.
The line labeled a) is the physical wavelength associated
with a fixed comoving scale $k$. The line b) is the Hubble radius
or horizon in SBB cosmology. Note that at $t_{\rm rec}$ the 
fluctuation is outside the Hubble radius. Curve c) shows the
Hubble radius during inflation. As depicted, at early
times during inflation the curve a) is inside the Hubble
radius, thus allowing for a causal generation mechanism for fluctuations
on the corresponding scale. The horizon in inflationary
cosmology is shown in curve d). The graph also demonstrates
the trans-Planckian problem of inflationary cosmology: at very
early times, the wavelength is smaller than the
Planck scale $\ell _{\rm Pl}$ (Phase I), at intermediate times
it is larger than $\ell _{\rm Pl}$ but smaller than the Hubble
radius $H^{-1}$ (Phase II), and at late times during
inflation it is larger than the Hubble radius (Phase III).
\label{Figone}}
\end{figure}

Let us briefly recall how inflationary cosmology leads to the existence
of a mechanism for using causal microphysics for producing fluctuations
on cosmological scales (when measured today), which at the time of equal
matter and radiation had a physical wavelength larger than the Hubble
radius $l_H(t) = H^{-1}(t)$, where $H = {\dot{a} / a}$ is the expansion
rate (for references to the original literature
see textbook treatments \cite{Linde,LiLy}). 
The relevant space-time sketch is shown in Fig. \ref{Figone}.
If we trace a fixed comoving length scale (corresponding e.g. to CMB
anisotropies on a fixed large angular scale) back into the past, then
in SBB cosmology this scale is larger than the horizon (which in
standard cosmology is $H^{-1}$ - up to a numerical coefficient of order one) 
at early times. The time when such scales ``cross''
the horizon is in fact later than $t_{\rm rec}$. Thus, it appears
impossible to explain the origin of the primordial fluctuations measured
in the CMB without violating causality \footnote{Topological defect models
are a counterexample to this ``standard dogma''}. 

In an inflationary
Universe (taking exponential inflation to be specific), 
during the period of exponential expansion the physical
wavelength corresponding to a fixed comoving scale decreases exponentially
as we go back in time, whereas the Hubble radius $l_H(t)$ is
constant \footnote{The Hubble radius
measures the distance over which microphysics can act coherently
at a fixed time $t$, and in inflationary cosmology differs in
a crucial way from the causal horizon (the forward light cone, which becomes
exponentially larger).}. Thus, as long as the period of inflation is
sufficiently long, all scales of cosmological interest today originate
inside the Hubble radius in the early stages of inflation. As will be
shown in the next section, the squeezing of quantum vacuum fluctuations
when they exit and propagate outside the Hubble radius is the mechanism
for the origin of fluctuations in inflationary cosmology. 

The same background dynamics which yields the causal generation mechanism
for cosmological fluctuations, the most spectacular success of inflationary
cosmology, bears in it the nucleus of the trans-Planckian problem. This
can also be seen from Fig. \ref{Figone}. If inflation lasts only slightly
longer than the minimal time it needs to last in order to solve the
horizon problem and to provide a
causal generation mechanism for CMB fluctuations, then the 
corresponding physical wavelength of these fluctuations
is smaller than the Planck length at the beginning of the period of inflation.
The theory of cosmological perturbations is based on classical general
relativity coupled to a weakly coupled scalar field description of
matter. Both the theory of gravity and of matter will break down
on trans-Planckian scales, and this immediately leads to the trans-Planckian
problem: are the predictions of standard inflationary cosmology robust
against effects of trans-Planckian physics \cite{RHBrev}?
To answer this question, we need to give the reader a short review of
the standard theory of cosmological fluctuations.

\section{Fluctuations in Inflationary Cosmology}

In the following we give an overview of the quantum theory of
cosmological perturbations (see review articles \cite{MFB} 
for details and references to the original literature).  
Since gravity is a purely attractive force, and since
the fluctuations on scales of the CMB anisotropies were small in amplitude
when the
anisotropies were generated, the fluctuations had to have been very
small in the early Universe. Thus, a linearized analysis of the
fluctuations is justified. In this case, the Fourier modes of the
cosmological fluctuations evolve independently.  

The basic idea of the theory of cosmological perturbations 
is to quantize the linear
fluctuations about a classical background cosmology.
The starting point is the full
action of gravity plus matter 
\begin{equation}
S \, = \, \int {\rm d}^4x \sqrt{-g}R +S_{\rm m} , 
\end{equation}
where the first term is the usual Einstein-Hilbert action for gravity,
$R$ being the Ricci scalar and $g$ the determinant of the metric, and
$S_{\rm m}$ is the matter action. For the sake of simplicity, and
since it is the usual assumption in simple inflationary Universe
models, we take matter to be described by a single minimally coupled
scalar field $\varphi$. Then, we separate the metric and matter into
classical background variables $g^{(0)}_{\mu \nu}, \varphi^{(0)}$
which depend only on time, and fluctuating fields $\delta g_{\mu \nu},
\delta\varphi$ which depend on space and time and have vanishing
spatial average:
\begin{equation}
g_{\mu \nu} \, =\, g^{(0)}_{\mu \nu}(\eta) +
\delta g_{\mu \nu}(\eta ,{\bf x}), 
\quad 
\varphi \, =\,\varphi^{(0)}(\eta ) +
\delta\varphi (\eta ,{\bf x})\, .  
\end{equation}
We will focus on {\it scalar metric fluctuations}, the fluctuating
degrees of freedom which couple to matter perturbations \footnote{Greek
variables run over space-time indices, Latin variables only over spatial
indices.}.

The description of
scalar metric perturbations is complicated by the fact that 
some perturbation modes correspond to space-time
reparametrizations of a homogeneous and isotropic cosmology. 
A simple way to address this issue of gauge is to
work in a system of coordinates which completely fixes the gauge. A
simple choice is the {\it longitudinal} gauge, in which the metric takes
the form \cite{MFB}
\begin{equation}
{\rm d}s^2 \, = \, a^2(\eta) \bigl[(1 + 2 \Phi) d\eta^2 - 
(1 - 2 \Psi) \delta_{i j}
{\rm d}x^i {\rm d}x^j \bigr] \, ,
\end{equation}
where the space- and time-dependent functions $\Phi$ and $\Psi$ are the
two physical metric degrees of freedom which describe scalar metric
fluctuations, and $\eta$ is conformal time (related to physical
time $t$ via $dt = a(\eta) d\eta$). 
The fluctuations of matter fields give additional degrees
of freedom for scalar metric fluctuations. In the simple case of a
single scalar matter field, the matter field fluctuation 
is $\delta \varphi$.  

In the absence of anisotropic stress, it
follows from the $i \neq j$ Einstein equations that the two metric fluctuation
variables $\Phi$ and $\Psi$ coincide. Due to the Einstein constraint
equation, the remaining metric fluctuation $\Psi$ is determined by the
matter fluctuation $\delta \varphi$. It is clear from this analysis of
the physical degrees of freedom that the action for scalar metric
fluctuations must be expressible in terms of the action of a single
free scalar field $v$ with a time dependent mass (determined by the
background cosmology). As shown in \cite{M85} (see also \cite{L80}),
this field is
\begin{equation}
v \, =\, a \biggl( \delta \varphi + {{\varphi^{(0)'}} 
\over {\ch}} \Psi \biggr) = z \calr \, ,
\end{equation}
where  
$\ch = a'/ a$ (a prime denoting the derivative with respect to $\eta$), 
\begin{equation}\label{zdef}
z \, \equiv \, a {{\varphi^{(0)'}} \over {\ch}} \, ,
\end{equation}
and $\calr$ denotes the curvature perturbation in comoving 
gauge \cite{L85}. The action for scalar metric fluctuations is 
\cite{M88}
\begin{equation}\label{scaleac}
\delta S_{\calr} \, =\, {1 \over 2} \int {\rm d}^4{\bf x} 
\biggl[(v_{\bf k}')^2- \delta^{i j} 
v_{{\bf k}, i} v_{{\bf k}, j} + {{z''} \over z} v_{\bf k}^2 
\biggr] \, ,
\end{equation}
which leads to the equation of motion
\begin{equation}\label{scaleeq}
v_{\bf k}'' + \biggl( k^2 - {{z''} \over z} \biggr) v_{\bf k} \, 
= \, 0 \, ,
\end{equation}
the equation of motion of a harmonic oscillator with time-dependent
mass, the time dependence being given by $z(\eta)$ which is a function
of the background cosmology. Note that if $a(\eta)$
is a power of $\eta$, then $\varphi_0'$ and $\ch$ scale with the
same power of $\eta$, and the variable $z$ is then proportional to $a$.

At this point we can summarize the mechanism by which cosmological
fluctuations are generated in inflationary cosmology: we canonically
quantize the linearized metric/matter fluctuations about a classical
background cosmology, solve the resulting dynamical problem as
a standard initial value problem, setting off all Fourier modes of the
scalar field $v$ in their vacuum state at the initial time (e.g.
the beginning of the period of inflation). 

The equation (\ref{scaleeq}) is a harmonic oscillator equation
with a time-dependent mass given by $z''/z$. On scales
smaller than the Hubble radius ($t<t_{\rm H}(k)$), 
the mass term is negligible, and the
mode functions oscillate with constant amplitude. On scales larger
than the Hubble radius, however, the mass term dominates and the $k^2$
term can be neglected. The mode functions no longer oscillate. In an
expanding background, the dominant mode of 
$v_{\bf k}(\eta)$ scales as $z(\eta)$.  Note that it is
incorrect to assume that fluctuations are created at the time of
Hubble radius crossing (an impression one could get from reading
\cite{Shenker}). Rather, the role of the time $t_{\rm H}(k)$
of Hubble radius crossing is to set the time when the classical mode
functions cease to oscillate and
begin to increase in amplitude (squeezing).

The quantum mechanical interpretation of the two phases $t<t_{\rm
H}(k)$ and $t>t_{\rm H}(k)$ is the following: on sub-Hubble scales we
have oscillating quantum vacuum fluctuations and there is no 
particle production. Once the scales cross the Hubble radius, the mode
functions begin to grow and the fluctuations get frozen. The
initial vacuum state then becomes highly squeezed for $t\gg t_{\rm H}(k)$
\footnote{The equation of motion for gravitational waves in an
expanding background cosmology is identical to (\ref{scaleeq}) with
$z(\eta)$ replaced by $a(\eta)$, and thus the physics is identical.
The case of gravitational waves 
was first discussed in \cite{Grishchuk}.}.  The squeezing leads to
the generation of effectively classical cosmological perturbations.

For cosmological applications, it is particularly interesting to
calculate the power spectrum of the curvature perturbation
$\calr$, defined as
\begin{equation}
\label{scalepow}
{\calp}_{\calr}(k) \, 
= \, {{k^3} \over {2 \pi^2}z^2}\vert v_k \vert ^2\, .
\end{equation}
This last quantity can be estimated very easily. From the fact that 
on scales larger than the Hubble radius the mode
functions are proportional to $a(\eta)$, we find
\begin{equation}
\label{final}
{\calp}_{\calr}(k) \sim {{k^3} \over {2 \pi^2}} {1 \over 2k}
\frac{1}{a^2[\eta_{\rm H}(k)]} \, ,
\end{equation}
where $\eta_{\rm H}(k)$ is the conformal time of Hubble radius crossing for
the mode with comoving wavenumber $k$. Note that the second factor on
the r.h.s. of (\ref{final}) represents the vacuum normalization of the
wave function.

As is evident from Figure 1, the standard theory of cosmological
fluctuations summarized in this section relies on extrapolating
classical general relativity and weakly coupled scalar matter field
theory to length scales smaller than the Planck length. Thus, it is 
legitimate to ask whether the predictions resulting from this theory
are sensitive to modifications of physics on physical length scales
smaller than the Planck length. There are various ways in which such
physics could lead to deviations from the standard predictions. First,
new physics could lead to a non-standard evolution of the initial
vacuum state of fluctuations in Period I, such that at Hubble radius
crossing the state is different from the vacuum state. A model of
realizing this scenario is summarized in the following section.
Secondly, trans-Planckian physics may lead to different boundary
conditions, thus resulting in a different final state. A string-motivated
example for this scenario is given in Section 4. 

\section{Trans-Planckian Analysis I: Modified Dispersion Relations}

The simplest way of modeling the possible effects of trans-Planckian
physics, while keeping the mathematical analysis simple, is to replace
the linear dispersion relation $\omega _{_{\rm phys}}=k_{\rm phys}$
of the usual equation for cosmological perturbations by
a non standard dispersion relation $\omega _{_{\rm phys}}=\omega
_{_{\bf phys}}(k)$ which differs from the standard one only
for physical wavenumbers larger than the Planck scale. This method
was introduced \cite{Unruh,CJ} in the context of studying the dependence of
thermal spectrum of black hole radiation on trans-Planckian physics. 
In the context of cosmology, it has been shown \cite{MB,BM,Niemeyer} 
that this amounts to replacing $k^2$ appearing in
(\ref{scaleeq}) with $k_{\rm eff}^2(n,\eta )$ defined by
\begin{equation}
k^2 \, \rightarrow \, k_{\rm eff}^2(k,\eta ) \equiv 
a^2(\eta )\omega _{_{\rm phys}}^2\biggl[\frac{k}{a(\eta )}\biggr].
\end{equation}
For a fixed comoving mode, this implies that the dispersion relation
becomes time-dependent. Therefore, the equation of motion of the
quantity $v_k(\eta)$ takes the form
\begin{equation} 
\label{eom2}
v_k'' + \biggl[k_{\rm eff}^2(k,\eta ) - {{a''} 
\over a}\biggr]v_k \, = \, 0 \, .
\end{equation}
A more rigorous derivation of this 
equation, based on a variational principle, has been provided \cite{LLMU} 
(see also Ref.~\cite{jacobson}).

The evolution of modes thus must be considered separately in three
phases, see Fig.~\ref{Figone}. In Phase I the wavelength is smaller
than the Planck scale, and trans-Planckian physics can play
an important role. In Phase II, the wavelength is larger than the
Planck scale but smaller than the Hubble radius. In this phase,
trans-Planckian physics will have a negligible effect
(this statement can be quantified \cite{Shenker}). Hence,
by the analysis of the previous section, the wave function of fluctuations is
oscillating in this phase, 
\begin{equation}
\label{vsubH}
v_k \, = \, B_1\exp(-ik\eta )+B_2\exp(ik\eta )
\end{equation}
with constant coefficients $B_1$ and $B_2$. In the standard approach,
the initial conditions are fixed in this region and the usual choice
of the vacuum state leads to $B_1=1/\sqrt{2k}$, $B_2=0$.  
Phase III starts at the time $t_{\rm H}(k)$ when the
mode crosses the Hubble radius. During this phase, the wave function is
squeezed.

One source of trans-Planckian effects \cite{MB,BM} on observations
is the possible non-adiabatic evolution of the wave function
during Phase I. If this occurs, then it is possible that the
wave function of the fluctuation mode is not in its vacuum state when
it enters Phase II and, as a consequence, the coefficients $B_1$ and
$B_2$ are no longer given by the standard expressions above. In this
case, the wave function will not be in its vacuum state when it crosses
the Hubble radius, and the final spectrum will be different. In
general, $B_1$ and $B_2$ are determined by the matching conditions
between Phase I and II. By focusing only \cite{Shenker} on
trans-Planckian effects on the local vacuum wave function at the
time $t_{\rm H}(k)$, one misses this important
potential source of trans-Planckian signals in the CMB.
If the dynamics is adiabatic
throughout (in particular if the $a''/a$ term is negligible), the WKB
approximation holds and the solution is always given by
\begin{equation} 
\label{WKBsol}
v_k (\eta )\, \simeq \, \frac{1}{\sqrt{2k_{\rm eff}(k,\eta )}}
\exp\biggl[-i\int _{\eta _{\rm i}}^{\eta }k_{\rm eff}{\rm d}\tau \biggr]
\, ,
\end{equation} 
where $\eta_i$ is some initial time. Therefore, if we start with
a positive frequency solution only and use this solution, we find
that no negative frequency solution appears. Deep in Region II where
$k_{\rm eff} \simeq k$ the solution becomes
\begin{equation}
v_k(\eta ) \simeq {1 \over {\sqrt{2k}}} \exp(-i \phi - i k \eta),
\end{equation}
i.e. the standard vacuum solution times a phase which will disappear
when we calculate the modulus. To obtain 
a modification of the inflationary spectrum, it is sufficient to 
find a dispersion relation such that the WKB approximation breaks down 
in Phase I.

A concrete class of dispersion relations
for which the WKB approximation breaks down is
\begin{equation}
\label{disprel}
k_{\rm eff}^2(k,\eta ) = k^2 - k^2 \vert b_m \vert
\biggl[{{\ell_{_{\rm pl}}} \over {\lambda(\eta)}} \biggr]^{2m}, 
\end{equation}
where $\lambda (\eta )=2\pi a(\eta )/k$ is the wavelength of a
mode. If we follow the evolution of the modes in Phases I, II and
III, matching the mode functions and their derivatives at the
junction times, the calculation \cite{MB,BM,MB2}
demonstrates that the final spectral index is modified
and that superimposed oscillations appear. 

However, the above example
suffers from several problems. First, in inflationary models with a
long period of inflationary expansion, 
the dispersion relation (\ref{disprel}) leads to complex
frequencies at the beginning of inflation for scales which are of current 
interest in cosmology. Furthermore, the initial conditions for the
Fourier modes of the fluctuation field have to be set in a region
where the evolution is non-adiabatic and the use of the usual vacuum
prescription can be questioned. These problems can be avoided in a toy
model in which we follow the evolution of fluctuations in a bouncing
cosmological background which is asymptotically flat in the past
and in the future. The analysis \cite{MB3} shows that even in this
case the final spectrum of fluctuations depends on the specific 
dispersion relation used.

An example of a dispersion relation which breaks the WKB approximation in
the trans-Planckian regime but does not lead to the problems mentioned in
the previous paragraph was investigated in \cite{LLMU}. It is a
dispersion relation which is linear for both small and large wavenumbers, but
has an intermediate interval during which the frequency decreases as
the wavenumber increases, much like what happens in (\ref{disprel}).
The violation of the WKB condition occurs for wavenumbers near the local 
minimum of the $\omega(k)$ curve.

\section{Trans-Planckian Analysis II: Space-Time Uncertainty Relation}

A justified criticism against the method summarized in the previous section
is that the non-standard dispersion relations used are completely ad hoc,
without a clear basis in trans-Planckian physics. There has been a lot
of recent work \cite{EG1,KN,EG2,Mangano} on the implication of space-space
uncertainty relations \cite{Ven,Gross} on the evolution of fluctuations.
The application of the uncertainty relations on the fluctuations lead to
two effects \cite{Kempf,Hassan}. 
Firstly, the equation of motion of the fluctuations
in modified. Secondly, for fixed comoving length scale $k$, the uncertainty
relation is saturated before a critical time $t_i(k)$. Thus, in addition
to a modification of the evolution, trans-Planckian physics leads to a
modification of the boundary condition for the fluctuation modes. The
upshot of this work is that the spectrum of fluctuations is modified.

We have recently studied \cite{Ho} the implications of the stringy
space-time uncertainty relation \cite{Yoneya,MY} 
\begin{equation}
\Delta x_{\rm phys} \Delta t \, \geq \, l_s^2 
\end{equation}
on the spectrum of cosmological fluctuations. Again, application of this
uncertainty relation to the fluctuations leads to two effects. Firstly,
the coupling between the background and the fluctuations is nonlocal in
time, thus leading to a modified dynamical equation of motion. Secondly,
the uncertainty relation is saturated at the time $t_i(k)$ when the physical
wavelength equals the string scale $l_s$. Before that time it does not
make sense to talk about fluctuations on that scale. By continuity,
it makes sense to assume that fluctuations on scale $k$ are created at
time $t_i(k)$ in the local vacuum state (the instantaneous WKB vacuum
state).

Let us for the moment neglect the nonlocal coupling
between background and fluctuation, and thus consider the usual
equation of motion for fluctuations in an accelerating background
cosmology.  We distinguish two ranges of scales. 
Ultraviolet modes are generated at late times when the Hubble radius is
larger than $l_s$. On these scales, the spectrum of fluctuations does not
differ from what is predicted by the standard theory, since at the time
of Hubble radius crossing the fluctuation mode will be in its vacuum
state. However, the evolution of infrared modes which are created when
the Hubble radius is smaller than $l_s$ is different. The fluctuations
undergo {\it less} squeezing than they do in the absence of the
uncertainty relation, and hence the final amplitude of fluctuations is
lower. From the equation (\ref{final}) for the power spectrum of
fluctuations, and making use of the condition
\begin{equation}
a(t_i(k)) \, = \, k l_s 
\end{equation}
for the time $t_i(k)$ when the mode is generated, it follows immediately
that the power spectrum is scale-invariant
\begin{equation} \label{finalb}
{\calp}_{\calr}(k) \, \sim \, k^0 \, . 
\end{equation}
In the standard scenario
of power-law inflation the spectrum is red (${\calp}_{\calr}(k) \sim k^{n-1}$
with $n < 1$). Taking into account the
effects of the nonlocal coupling between background and fluctuation
mode leads \cite{Ho} to a modification of this result: the spectrum
of fluctuations in a power-law inflationary background is in fact blue 
($n > 1$). 

Note that, if we neglect the nonlocal coupling between background and
fluctuation mode, the result of (\ref{finalb}) also holds in a 
cosmological background which is NOT accelerating. Thus, we have a method
of obtaining a scale-invariant spectrum of fluctuations without inflation.
This result has also been obtained in \cite{Wald}, however without a
microphysical basis for the prescription for the initial conditions.

\section{Non-Commutative Inflation}

A key problem with the method of modified dispersion relations
is the issue of back-reaction \cite{Tanaka,Starob}. If the mode
occupation numbers of the fluctuations at Hubble radius crossing are
significant, the danger arises that the back-reaction of the fluctuations
will in fact prevent inflation. Another constraint arises
from the observational limits on the flux of ultra-high-energy
cosmic rays. Such cosmic rays would be produced \cite{Tkachev} in
the present Universe if Trans-Planckian effects of the type
discussed in the two previous sections were present. These issues 
are currently under investigation.

Surprisingly, it has been realized \cite{Stephon}
that back-reaction effects due to
modified dispersion relations (which in turn are motivated by string theory)
might it fact yield a method of obtaining inflation from pure radiation.
In this section, $p$ and $\omega$ will denote physical wavenumber and
frequency, respectively.

One of the key features expected from string theory is the existence
of a minimal length, or equivalently a maximal wavenumber $p_{\rm max}$. 
Thus, if we consider the dispersion relation for pure radiation in string 
theory, the dispersion relation must saturate or turn over at $p_{\rm max}$.
Some cosmological consequences of a dispersion relation which saturates
at $p_{\rm max}$, i.e. for which the frequency $\omega$ diverges as
$k \rightarrow p_{\rm max}$ were explored recently \cite{AlMa}. It was
found that a realization of the varying speed of light scenario can be
obtained.

Recent work \cite{Stephon} has focused on the case when the dispersion
relation turns over at $p_{\rm max}$, i.e. if the frequency is increased, 
then before $p(\omega)$ reaches $p_{\rm max}$, the
wavenumber begins to decrease again. This implies that for each value
of $p$ there are two states corresponding to two different frequencies,
i.e. that there are two branches of the dispersion relation. A class of
such dispersion relations is given by
\begin{equation} \label{disprelb}
\omega^2 - p^2 f^2 \, = \, 0 \,\,\, , \, f(\omega) \, = \, 1 + 
(\lambda E)^{\alpha} \, ,
\end{equation}
where $\alpha$ is a free parameter.

Let us now consider an expanding Universe with such a dispersion relation, and
assume that in the initial state both branches are populated up to a
frequency much larger than the frequency at which the dispersion relation
turns over. As the Universe expands, the physical wavenumber of all modes
decreases. However, this implies - in contrast to the usual situation -
that the energy of the upper branch states {\it increases}. This is what
one wants to be able to achieve an inflationary cosmology. Eventually,
the high energy states will decay into the lower branch states (which
have the usual equation of state of radiation), thus leading to a
graceful exit from inflation.

To check whether the above heuristic scenario is indeed realized, one
must compute the equation of state corresponding to the dispersion
relation (\ref{disprelb}). The spectrum is deformed, and a thermodynamic
calculation \cite{Stephon} yields an equation of state which in the
high density limit tends to
%\begin{equation}
$P  = 1 / [3(1 - \alpha)] \rho$
%\end{equation}
where $P$ stands for the pressure. Thus, we see that there is a narrow
range of values of $\alpha$ which indeed give the correct equation of
state for power-law inflation.

In such an inflationary scenario, the fluctuations are of thermal origin.
Taking the initial r.m.s. amplitude of the mass fluctuations on thermal
length scale $T^{-1}$ to be order unity, assuming random superposition
of such fluctuations on larger scales to compute the amplitude of
fluctuations when a particular comoving length scale exits the Hubble
radius, and using the usual theory of cosmological fluctuations to track
the fluctuations to the time when the scale enters the Hubble radius,
one finds a spectrum of fluctuations with the same slope as
in regular power-law inflation, and with an amplitude
which agrees with the value required from observations if the fundamental
length scale $l_s$ at which the dispersion relation turns over is about an 
order of magnitude larger than the Planck length.

\section{Conclusions}

Due to the exponential redshifting of wavelengths, present cosmological
scales originate at wavelengths smaller than the Planck length early on
during the period of inflation. Thus, Planck physics may well encode
information in these modes which can now be observed in the spectrum of
microwave anisotropies. Two examples have been shown to demonstrate the
existence of this ``window of opportunity'' to probe trans-Planckian
physics in cosmological observations. The first method makes use of
modified dispersion relations to probe the robustness of the predictions
of inflationary cosmology, the second applies the stringy space-time
uncertainty relation on the fluctuation modes. Both methods yield the
result that trans-Planckian physics may lead to measurable effects in
cosmological observables. An important issue which must be studied more
carefully is the back-reaction of the cosmological fluctuations (see e.g.
\cite{ABM} for a possible formalism). As demonstrated in the final section,
it is possible that trans-Planckian physics can in fact lead to dramatic
changes even in the background cosmology.

A final remark related to other work on the trans-Planckian problem for
inflationary cosmology: in our view, nontrivial trans-Planckian physics
(which cannot be described in any way by a free scalar field) produces
excitations about the Bunch-Davies vacuum state for a finite range of
wavelengths. On these wavelengths, the excited state may (but need not)
look like a nontrivial $\alpha$ vacuum of a free scalar field theory
in de Sitter space \cite{UD1,UD2,Easther3,Lowe}. However, globally
the state is NOT an $\alpha$ vacuum, so that the recent discussions 
about possible problems with $\alpha$ vacua \cite{Banks,Einhorn,KKLSS,UD3}
are not crucial to the question of whether Planck-scale physics can 
lead to distinctive signatures observable in cosmology.

\section*{Acknowledgments}

This is a slightly updated version of the contribution to the
proceedings of the CosPA2002 meeting at National Taiwan University
held May 31 - June 2 2002.
I am grateful to the organizers of CosPA2002 for their wonderful
hospitality, and to S. Alexander, P.-M. Ho, S. Joras,
J. Magueijo and J. Martin for collaboration.
The research was supported in part
by the U.S. Department of Energy under Contract DE-FG02-91ER40688, TASK A.

\end{document}